\documentclass[a4paper]{article}

\usepackage[english]{babel}
\usepackage[utf8x]{inputenc}
\usepackage[T1]{fontenc}
\usepackage{authblk}

\usepackage[a4paper,top=4cm,bottom=3cm,left=4cm,right=4cm,marginparwidth=1.75cm]{geometry}

\usepackage{multirow}

\usepackage[round]{natbib}

\usepackage{amsmath}
\usepackage{amssymb}
\usepackage{graphicx}
\usepackage{subcaption}
\usepackage[colorinlistoftodos]{todonotes}
\usepackage[colorlinks=true, allcolors=blue]{hyperref}

\usepackage{float}
\usepackage{bm}
\label{key}
\usepackage{color}

\makeatletter
\def\blfootnote{\xdef\@thefnmark{}\@footnotetext}
\makeatother

\title{Interpreting the Ising Model: The Input Matters}

\author{Jonas M. B. Haslbeck}
\author{Sacha Epskamp}
\author{Maarten Marsman}
\author{Lourens J. Waldorp}

\affil{Psychological Methods Group, University of Amsterdam}

\date{}

\begin{document}
	\maketitle
	
	\begin{abstract}
	The Ising model is a model for pairwise interactions between binary variables that has become popular in the psychological sciences. It has been first introduced as a theoretical model for the alignment between positive (1) and negative (-1) atom spins. In many psychological applications, however, the Ising model is defined on the domain $\{0,1\}$ instead of the classical domain $\{-1,1\}$. While it is possible to transform the parameters of the Ising model in one domain to obtain a statistically equivalent model in the other domain, the parameters in the two versions of the Ising model lend themselves to different interpretations and imply different dynamics, when studying the Ising model as a dynamical system. In this tutorial paper, we provide an accessible discussion of the interpretation of threshold and interaction parameters in the two domains and show how the dynamics of the Ising model depends on the choice of domain. Finally, we provide a transformation that allows one to transform the parameters in an Ising model in one domain into a statistically equivalent Ising model in the other domain.

	\end{abstract}
	
	\blfootnote{This article has been accepted for publication in Multivariate Behavioral Research,
		published by Taylor \& Francis.}
	
	
\section{Introduction}\label{sec_intro}

The Ising model is a model for pairwise interactions between binary variables that originated in statistical mechanics \citep{ising1925beitrag, glauber1963time} but is now used in a large array of applications in the psychological sciences \citep[e.g.][]{borsboom2013network, marsman2015bayesian, boschloo2015network, boschloo2016network, fried2015loss, cramer2016major, dalege2016toward, rhemtulla2016network, van2017network, haslbeck2017predictable, afzali2017network, deserno2017multicausal, savi2018wiring, marsman2019network} 

The original Ising model has been introduced as a model for the interactions between atom spins, which can be positive (1) and negative (-1)  \citep{brush1967history}. In this setting, with variables taking values in the domain $\{-1,1\}$, the interaction parameters in the Ising model determine the \emph{alignment} between variables: If an interaction parameter between two variables is positive, the two variables tend to take on the same value; on the other hand, if the interaction parameter is negative, the two variables tend to take on different values. 

In most psychological applications, however, the Ising model is defined with variables taking values in the domain $\{0,1\}$. While it is possible to transform the parameters of a given Ising model in one domain to obtain a statistically equivalent model in the other domain, the parameters in the two versions of the Ising model lend themselves to different interpretations and imply different dynamics, when studying the Ising model as a dynamical system. If unaware of those subtle differences, one might erroneously apply theoretical results from the $\{-1,1\}$ domain to an estimated model in the $\{0,1\}$ domain, or simply interpret parameters incorrectly. To prevent such confusion in the emerging psychological networks literature which makes heavy use of the Ising model, we provide a detailed discussion of both versions of the Ising model in the present tutorial paper.

We begin by discussing the different interpretations of the Ising model in the $\{-1,1\}$ and $\{0,1\}$ domain in Section \ref{sec_dd_p2}, using a simple example with two variables which allows the reader to follow all calculations while reading. We explain the differences in the interpretation of the threshold and interaction parameters in the two versions of the Ising model, and discuss in which situation which version might be more appropriate. While most psychological applications of the Ising model use it as a statistical model, it has also been studied as a dynamical system in psychological research \citep[e.g.,][]{cramer2016major, dalege2016toward, lunansky2019personality}. In Section \ref{sec_con_dynamics} we discuss how the dynamics of the Ising model depends on the choice of domain, and show that the domain changes the \emph{qualitative} behavior of the model. Finally, in Section \ref{sec_IsingTrans_short} we provide a transformation that allows one to transform the parameters in an Ising model in one domain into a statistically equivalent Ising model in the other domain.

\section{Different Domain, Different Interpretation}\label{sec_dd_p2}
	
In this section we estimate an Ising model with $p=2$ variables in both domains, $\{-1, 1 \}$ and $\{0, 1\}$, and show that the resulting threshold and interaction parameters have different values and lend themselves to different interpretations. We choose the  $p=2$ variable case to make the explanation as accessible as possible. However, all results immediately extend to the general situation with $p$ variables. The Ising model for two variables is given by
	
\begin{equation}\label{twovar_Ising}
P(y_1, y_2) = \frac{1}{Z} \exp \left \{ \alpha_1 y_1 + \alpha_2 y_2 + \beta_{12} y_1 y_2 \right \},
\end{equation}
	
	\noindent
	where $y_1, y_2$ are either elements of $\{-1,1\}$ or $\{0,1\}$,  $P(y_1, y_2)$ is the probability of the event $(y_1, y_2)$, $\alpha_1, \alpha_2, \beta_{12}$ are parameters in $\mathbb{R}$, and $Z$ is a normalization constant which denotes the sum of the exponent across all possible states. There are $2^p=4$ states in this example.
	
	To illustrate the differences across models, we generate $n = 1000$ samples of the labels $A, B$ with the relative frequencies shown in Table \ref{table_datagen}:
	
	\begin{table}[H]
		$$
		\bordermatrix{~ & A & B \cr
			A & 0.14 & 0.18 \cr
			B & 0.18 & 0.50 \cr}
		$$
		\caption{Relative frequency of states in the example data set.}\label{table_datagen}
	\end{table}
	
	In applications, the labels $A, B$ can stand for any pair of categories such as being for or against something, some event having happened or not, or a symptom being present or not. The two domains are two different ways to numerically represent these labels.
	
	We obtain the Maximum Likelihood Estimates (MLE) of the parameters in two different ways: once, by filling in $\{-1, 1 \}$ for $\{A, B\}$; and once by filling in $\{0, 1 \}$ for $\{A, B\}$. Figure \ref{fig_intro} summarizes the two resulting models:
	
	\begin{figure}[H]
		\centering
		\includegraphics[width=1\textwidth]{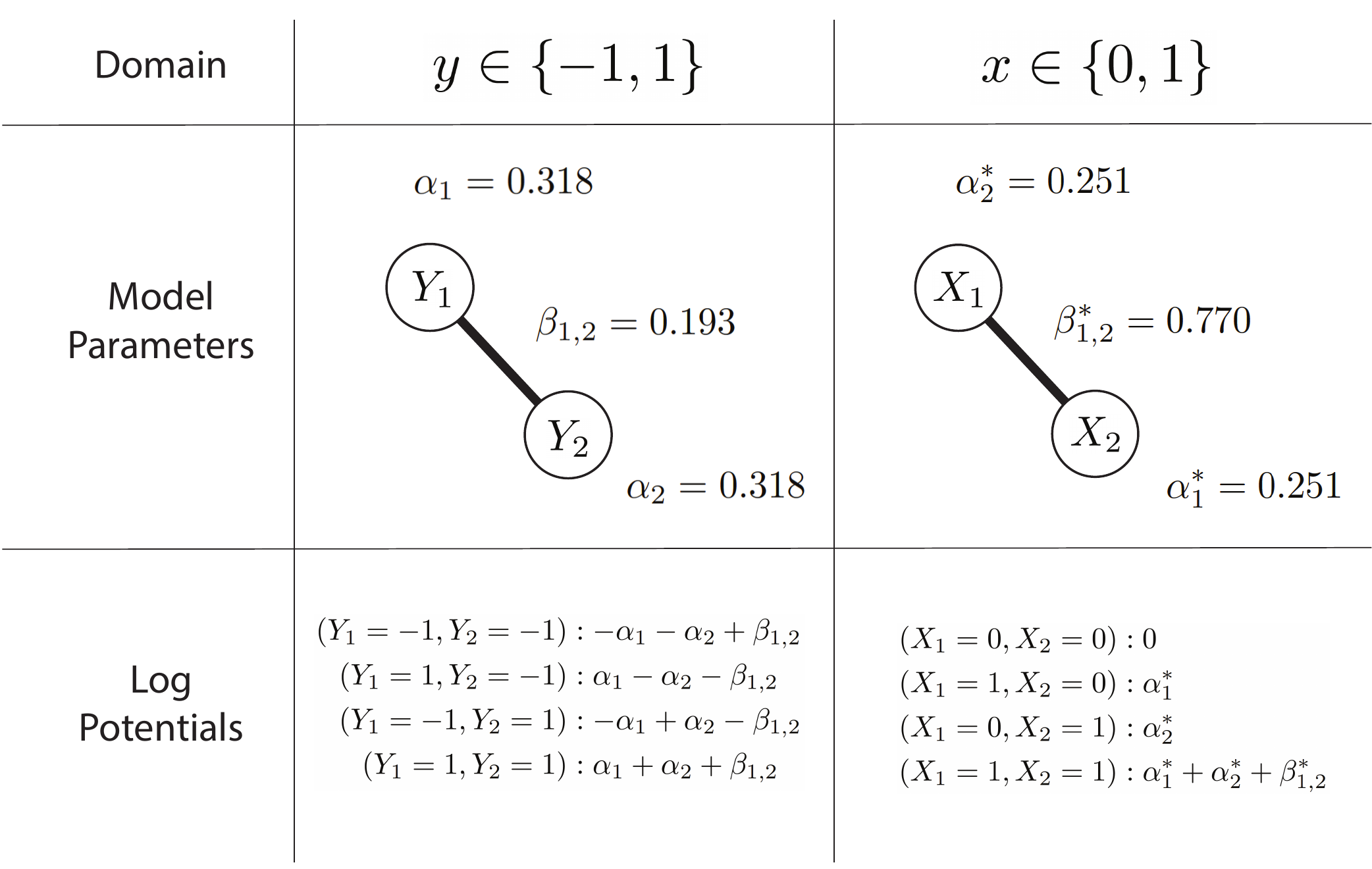}
		\caption{The threshold and interaction parameters estimated from the data generated from  Table \ref{table_datagen}, and the log potentials for each combination of states, separately for the two domains $\{-1, 1 \}$ and $\{0, 1 \}$. The log potentials (also called energy function or Hamiltonian) are obtained by filling each state (e.g., $y_1=-1, y_2=-1$) in the expression within the exponential in equation (\ref{twovar_Ising}).
		}\label{fig_intro}
	\end{figure}
	
	The first column in Figure \ref{fig_intro} shows the parameter estimates $\alpha_1, \alpha_2$ and $\beta_{12}$, and log potentials in domain $\{-1, 1 \}$. We first focus on the interpretation of the interaction parameter $\beta_{12}$. To understand the interpretation of this parameter we take a look at the log potentials for all four states $\{ (-1,-1), (-1,1), (1,-1), (1,1) \}$, which we obtain by plugging the four states into the expression within the exponential in equation (\ref{twovar_Ising}). The resulting log potentials are displayed in the second row in Figure \ref{fig_intro} and show us the following: if $\beta_{12}$ becomes larger, the probability of the states $(-1,-1), (1,1) $ increases relative to the probability of the states $(-1,1), (1,-1)$. This means that the interaction parameter determines the degree of \emph{alignment} of two variables. That is, if $\beta_{12}>0$ the \emph{same} labels align with each other, and if $\beta_{12}<0$ \emph{opposite} labels align with each other. In other words, $\beta_{12}$ models the probability of the states $(-1,-1), (1,1) $ relative to the probability of the states $(-1,1), (1,-1)$.
	
	This is not the case in the $\{0, 1 \}$ domain. The second column in Figure \ref{fig_intro} shows that the parameter estimates $\alpha_1^*, \alpha_1^*, \beta_{12}^*$ in domain $\{0, 1 \}$, and we see that they have different values than in the $\{-1, 1 \}$ domain. To understand why this is the case, we again look at the interpretation of the interaction parameter $\beta_{12}^*$ by inspecting the four log potentials. The key observation is that $\beta_{12}^*$ only appears in the log potential of the state $(1,1)$. What happens if $\beta_{12}^*$ increases? Then the probability of the state $(1,1)$ increases relative to the probability of all other states $(0,1),(1,0), (0,0)$. In other words, $\beta_{12}^*$ models the probability of state  $(1,1)$ relative to the probability of the states $(0,1),(1,0),(0,0)$.
	
	Next, we turn to the interpretation of the threshold parameters. If all interaction parameters are equal to zero, the threshold parameters in both domains indicate the tendency of a variable to be in one state or the other. That is, $\alpha, \alpha^* > 0$ implies a larger probability for the states $(1) \in \{0, 1 \}$, $(1) \in \{-1, 1 \}$ than for states $(0) \in \{0, 1 \}$, $(-1) \in \{-1, 1 \}$. If $\alpha, \alpha^* < 0$ the reverse is true, and if $\alpha, \alpha^* = 0$, the corresponding states have both probability $0.5$. However, in the general case in which interaction parameters are allowed to be nonzero, the interpretation depends on the domain: in the $\{-1, 1 \}$ domain the threshold parameter indicates the tendency of a variable averaged over all possible states of all other variables. In more formal terms, the threshold parameter of a given variable indicates the marginal mean of that variable. In contrast, the threshold in the $\{0, 1 \}$ domain indicates the tendency of a variable when all other variables are equal to zero. We return to the different interpretations of thresholds in Section \ref{sec_con_dynamics}, in which we discuss the dynamics of the Ising model.
	
	In this section we showed that depending on its domain, the parameters of the Ising model have different interpretations. What are the consequences for applied researchers? In terms of reporting, it is important to state which domain has been used such that the reported model can be re-used in the correct way: if someone reports a set of parameters estimated from in the $\{0, 1 \}$ domain, and a reader applies it to the $\{-1, 1 \}$ domain they will obtain the incorrect probabilities. Note that in order to use the model one also has to report the threshold parameters. Not reporting the threshold parameters is a common problem and irrespective of the issue discussed in this paper. The only situation in which the domain does not matter is if the only goal is to compare the relative size of interaction parameters since the relative size is the same in both domains (see Section \ref{sec_IsingTrans_short}).
	
	The second consequence is that researchers have to choose which version of the Ising model is more appropriate for the phenomenon at hand. The above clarified interpretations of the Ising model in its two different domains allow to take this decision. For example, the $\{-1, 1\}$ parameterization may be more plausible for labels that are not qualitatively different, but rather opposing each other in some way such as supporting or opposing a certain viewpoint, for example agreeing or disagreeing with a statement like ``Elections should be held every two years instead of every four years''. This also reflects the origin of the Ising model as a model for atom spins, which are either positive or negative. The parameterization implied by $\{0, 1\}$ could be more appropriate if the two labels are qualitatively different, such as the presence or absence of an event or a characteristic. Take psychiatric symptoms as an example: while it seems plausible that \emph{fatigue} leads to \emph{lack of concentration}, it is less clear whether the absence of \emph{fatigue} also leads to the increase of \emph{concentration}. In such a case, we can encode the possible belief that the absence of something cannot have an influence on something else by choosing the $\{0, 1\}$ domain. Importantly, the decision of which version to pick has to be based on information beyond the data, because the models are statistically equivalent and therefore cannot be distinguished by observational data. In Appendix \ref{A_probcalc} we prove this equivalence for the example shown in Figure  \ref{fig_intro}.

	While Ising models in psychological research are usually fit to cross-sectional data, one is typically interested in within-subjects dynamics. In this context, one is often interested in inferring from an estimated Ising model how to best intervene on the system. In the next section we will show how the dynamics of the Ising model depends on its domain, and that the different versions of the Ising model make different predictions for optimal interventions.

	\section{Different Domain, Different Dynamics}\label{sec_con_dynamics}
	
	The choice of domain also determines the dynamics of the Ising model, when studying it as a dynamical system describing within-person dynamics. The dynamical version of the Ising model is initialized by $p$ initial values at $t=1$, and then each variable at time $t$ is a function of all variables it is connected to via a nonzero interaction parameter at $t-1$ \footnote{Glauber dynamics \citep{glauber1963time} describe a different way to sample from a dynamic Ising model. The qualitative results presented in this section also hold for Glauber dynamics.}. An often studied characteristic in this model is how its behavior changes when the size of the interaction parameters increases. A typical behavior of interest is the number of variables in state $(1)$ \citep[e.g.][]{dalege2016toward, cramer2016major}.
	
	Which behavior would we expect in the two domains $\{-1,1\}$ and $\{0,1\}$? From the previous section we know that in domain $\{-1,1\}$, the interaction parameter $\beta_{ij}$ models the probability of states $\{ (-1,-1), (1,1)\}$ relative to the states $\{(-1,1), (1,-1)\}$. Now, when increasing all $\beta_{ij}$, connected variables become more synchronized, which means that all (connected) variables tend to be either \emph{all} in state $(-1)$ or $(1)$. In terms of number of variables in state $(1)$, we would therefore predict that the expected number of variables in state $(1)$ remains unchanged, because the states $(-1)$ and $(1)$ occur equally often in the aligned ($(-1,-1)$ and $(1,1)$) and not aligned ($(-1,1)$ and $(1,-1)$) states.  And second, we predict that the probability that at a given time point either all variables are in state  $(-1)$, or all variables are in state $(1)$, increases. The reason is that, in the $(-1,1)$ domain, the larger the interaction parameter, the stronger the alignment between variables. This second prediction implies that the variance of the number of states in $(1)$ increases.
	
	In the domain $\{0,1\}$, the interaction parameter $\beta_{ij}^*$ models the probability of the state $(1,1)$ relative to the remaining three states $\{(0,1), (1,0), (0,0)\}$. Now, when increasing $\beta_{ij}^*$, connected variables will have a higher probability to be all in state $1$. Importantly, the frequency of $1$s in the high probability state $(1,1)$ is higher than in the other three states. We therefore expect that the number of variables in state $(1)$ increases and that the probability that all variables are in state $(1)$ increases. The second prediction implies that the variance of the number of states in $(1)$ decreases.
	
\begin{figure}[h]
		\centering
		\includegraphics[width=1\textwidth]{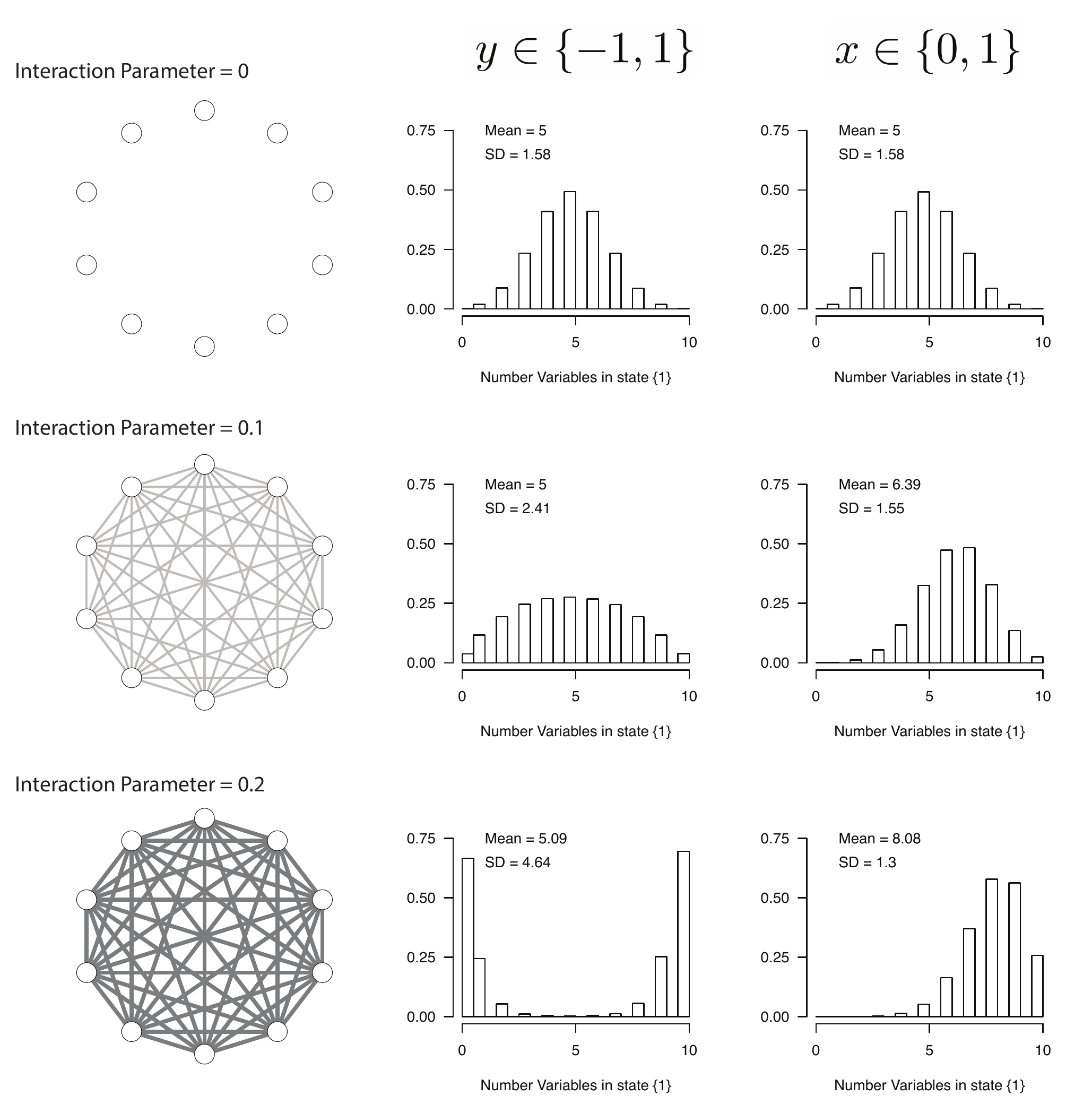}
		\caption{The distribution of the number of variables being in state one as a function of the size of the interaction parameter in a fully connected Ising model $\{0, .1, .2\}$ and used domain ($\{-1,1\}$ and $\{0,1\}$) of the random variable.}\label{fig_dynamics}
	\end{figure}

	We prove that the expected number of variables in state $(1)$ remains unchanged for $\{-1,1\}$ and increases for $\{0,1\}$, if $\beta_{ij} >0$ for the case $p=2$ variables in Appendix \ref{A_marginal_p2}. Here, we show via simulation that our predictions are correct. We sample $n = 10^6$ observations from a fully connected (i.e., all interaction parameters are nonzero) Ising model with $p=10$ variables in which all edge weights (interaction parameters) have the same size and all thresholds are set to zero. We vary both the size of the interaction parameters $\beta_{ij} \in \{0, .1, .2\}$ and the domain\footnote{The code to reproduce the simulation and Figure \ref{fig_dynamics} is available at \url{http://github.com/jmbh/IsingVersions}.}. Figure \ref{fig_dynamics} shows the distribution (over time steps) of the number of variables that are in state $\{1\}$.

	The first row of Figure \ref{fig_dynamics} shows the distribution of the number of variables in state $(1)$ across time when all interaction parameters are equal to zero. We see a symmetric, unimodal distribution with mean $5$ for both domains. This is what we would expect since the probability of each variable being in state $(1)$ can be seen as an unbiased (because the thresholds are zero) coin flip that is independent of all other variables. Thus, since we have 10 variables, the means are equal to $10 \times 0.5 = 5$. 
	
	 However, when increasing the interaction parameter from $0$ to $0.1$ (second row) the distributions become different: in domain $\{-1,1\}$ the mean remains unchanged and the probability mass shifts from around 5 to more extreme values, resulting in increased variance. In contrast, in domain $\{0,1\}$ the distribution shifts to the right, which implies that the mean increases and the variance slightly decreases. When further increasing the interaction parameters to $0.2$ (third row), in domain $\{-1,1\}$ most of the probability mass is concentrated on 0 and 10, while leaving the mean unchanged; in domain  $\{0,1\}$ the mean further increases and the variance further decreases. From a dynamical perspective, this means that for strongly connected Ising models (with thresholds equal to zero) the domain $\{-1,1\}$ implies two stable states (all variables in state $(-1)$ or $(1)$), while the domain $\{0,1\}$  implies only a single stable (all variables in state $(1)$), whose position depends on whether interaction parameters are positive or negative. This means that the dynamic Ising model in the $\{-1,1\}$ can switch between stable states, while $\{0,1\}$ it always stays in the same stable state\footnote{The result about bistability is true for the considered fully connected Ising model with zero thresholds. It is also possible to construct a bistable Ising model in the $\{0,1\}$ domain by choosing large negative thresholds and large positive interaction parameters. The relationship between mean/variance and changing the interaction parameter in the two domains, however, is always true.}.

	For the general case of Ising models that are not fully connected and also have negative interaction parameters, the results described above extend to local clusters of two or more variables: in the domain $\{-1,1\}$, increasing the interaction parameter will leave the means of all variables in the cluster unchanged, however, the variables become increasingly aligned (if interaction parameters are positive) or disaligned (if interaction parameters are negative). Alignment will lead to an increase in variance, while disalignment will lead to a decrease in variance. In contrast, in the $\{0,1\}$ domain the mean of all variables in the cluster will increase in the case of positive interaction parameters, and decrease in the case of negative interaction parameters.


	This shows that, depending on which domain is used one can come to entirely different conclusions about the dynamics of the Ising model. For example, \cite{cramer2016major} model the interactions between psychiatric symptoms with an Ising model in domain $\{0,1\}$ and conclude that densely connected Ising models imply a larger number of active (in state $(1)$) symptoms and therefore represent ``pathological'' models. The above argument and simulation show that this is only true when using the $\{0,1\}$ domain, which encodes the belief that the absence of a symptom cannot influence the absence of another symptom. If one decides that an alignment between variables is a more plausible interaction (as implied by the $\{-1,1\}$ domain), then densely connected Ising models do not imply a large number of active symptoms. Instead, high density implies high variance and two stable states. Thus, the characterization of dense networks as pathological networks as in \cite{cramer2016major} hinges on choosing the $\{0,1\}$ domain.
	
	This has important consequences: when choosing the $\{0,1\}$ domain, we would conclude that highly connected symptom networks are necessarily ``bad'', and interventions on the interactions between symptoms as suggested by \cite{borsboom2017network} should always reduce symptom activation. On the other hand, in the $\{-1,1\}$ domain highly connected symptom networks are not necessarily bad, but in fact can lead to high resilience, if the threshold parameters are large negative values. In such a situation strong interactions would keep the system in a state in which all symptoms are deactivated.

	\section{Transforming from $\{-1, 1\}$ to $\{0, 1\}$ and vice versa}\label{sec_IsingTrans_short}
	
	The Ising model is typically estimated by a sequence of $p$ logistic regressions, which require the domain $\{0, 1\}$. However, the previous sections showed that in some situations the domain $\{-1, 1 \}$ may be more appropriate. In Table \ref{tab:tab2_intro} we present a transformation that allows one to obtain the parameterization based on domain $\{-1, 1\}$ from the parameterization based on domain $\{0, 1\}$ and vice versa (see Appendix \ref{sec_IsingTrans} for the derivation of the transformations). We define $\beta_{i+}^\ast = \sum\limits_{j = i1}^{p} \beta_{ij}^\ast$ as the sum over the interaction parameters associated with a given variable $y_i$.
	
	\begin{table}[h]
		\centering
		\begin{tabular}{ r | c c }
			\textbf{Transformation} & Thresholds & Interactions \\
			\hline
			$\{0, 1\} \Rightarrow \{-1, 1\}$ & $\alpha_i = \frac{1}{2}\alpha_i^\ast + \frac{1}{4}\beta_{i+}^\ast$ & $\beta_{ij} = \frac{1}{4}\beta_{ij}^\ast$  \\
			$\{-1, 1\}  \Rightarrow \{0, 1\}$ & $\alpha_i^\ast = 2\alpha_i - 2 \beta_{i+}$ & $\beta_{ij}^\ast = 4\beta_{ij}$ 
		\end{tabular}
		\caption{Transformation functions to obtain the threshold and interaction parameters of one parameterization from the threshold and interaction parameters of the other parameterization. Parameters with asterisk indicate parameters in the $\{0, 1\}$ domain.}
		\label{tab:tab2_intro}
	\end{table}
	
	Table \ref{tab:tab2_intro} shows that the interaction parameters $\beta_{ij}$ in the $\{-1, 1\}$ domain are 4 times smaller than the interaction parameters $\beta_{ij}^*$ in the $\{0, 1\}$ domain. We also see that the threshold parameter $\alpha_i$ is a function of \emph{both} the threshold and the interaction parameters $\alpha_i^*, \beta_{ij}^*$ in the other parameterization.
	
	We now apply the transformations in Table \ref{tab:tab2_intro} to the $p=2$ variable example in Figure \ref{fig_intro}. We choose to transform from $\{0, 1\}$ to $\{-1, 1\}$:
	
	\begin{align*}
	a_1 & = \frac{1}{2} a_1^\ast + \frac{1}{4}\beta_{i+}^\ast =  \frac{0.251}{2}  + \frac{0.77}{4} = 0.318 \\
	\beta_{12} &= \frac{1}{4} \beta_{12}^\ast = \frac{.77}{4} = 0.1925 \approx 0.193 \\
	\end{align*}
	
	\noindent
	And indeed, we obtain the parameters obtained when estimating the Ising model in the $\{-1, 1 \}$ domain (see first column in Figure \ref{fig_intro}).
	
	From the transformation in Table \ref{tab:tab2_intro} follows that the two models are statistically equivalent. This implies that one could also estimate the model in the $\{-1, 1\}$ domain, transform the parameters, and would obtain the parameters one would have obtained from estimating in the $\{0, 1\}$ domain. Also, note that the standard errors of estimates are subject to the same transformation, and therefore one always reaches the same conclusion regarding statistical significance in both domains.
	
	However, note that one does not necessarily arrive at statistical equivalent models when estimating in the two different domains using biased estimators. An example of a biased estimation method is the popular $\ell_1$-regularized estimator \citep{van2014new}. We discuss why statistical equivalence is not guaranteed in this specific example in Appendix \ref{A_penMS}. The possibility that different domains lead to models that are not statistically equivalent highlights the importance of choosing the most plausible Ising model on substantive grounds.
	

	\section{Conclusions}
	
	In this paper we have investigated the subtleties in choosing the domain of the Ising model. We showed that estimating the Ising model in the domains $\{0, 1\}$ and $\{-1, 1\}$ results in parameters with different values and different interpretations. We also showed that the qualitative behavior of the dynamical Ising model depends on the chosen domain. Finally, we provided a transformation that explains the relation between the two parameterizations and allows one to obtain one from the other. This is useful in practice, because typically used software packages require the  $\{0, 1\}$ domain. This transformation also implies that the two parameterizations are statistically equivalent, which means that one cannot choose one over the other on empirical grounds. Thus, researchers should carefully reflect on which interactions between variables are plausible and choose the domain accordingly.

	\section*{Acknowledgements}
	
	 We would like to thank Joost Kruis, Ois\'in Ryan, Fabian Dablander, Jonas Dalege and Joris Broere for helpful comments on earlier versions of this paper. This project was supported by the European Research Council Consolidator Grant no. 647209 and NWO Veni Grant no. 451-17-017.

	\bibliographystyle{abbrvnat}
	\bibliography{IsingBib}
	
	\appendix
	
	\section{Statistical Equivalence worked out for two variable example}\label{A_probcalc}
	
	Here we show that the two models shown in Figure \ref{fig_intro} are statistically equivalent. Two models statistically equivalent if they output the same probability for any of states on which the models are defined.
	
	We begin with the model estimated on the domain $\{-1,-1\}$. We first compute the potentials for the four states $\{(-1,-1), (-1, 1), (1, -1), (1, 1)\}$:
	
	$$ 
	\exp \left \{
	0.318(-1) + 0.318(-1) + 0.193(-1)(-1)
	\right \} 
	= 0.6415304
	$$
	
	$$ 
	\exp \left \{
	0.318(-1) + 0.318(1) + 0.193(-1)(1)
	\right \} 
	= 0.8248249
	$$
	
	$$ 
	\exp \left \{
	0.318(1) + 0.318(-1) + 0.193(1)(-1)
	\right \} 
	= 0.8248249
	$$
	
	$$ 
	\exp \left \{
	0.318(1) + 0.318(1) + 0.193(1)(1)
	\right \} 
	= 2.29118
	$$
	
	\noindent
	and then the normalization constant 
	
	$$
	Z = 0.6415304+0.8248249+0.8248249+2.29118 = 4.58236
	$$
	
	\noindent
	We divide the  potentials by Z and obtain the probabilities
	
	$$
	P(Y_1 = -1, Y_2 = -1) = \frac{0.6415304}{Z} = 0.14
	$$
	
	$$
	P(Y_1 = -1, Y_2 = 1) = \frac{0.8248249}{Z} = 0.18
	$$
	
	$$
	P(Y_1 = 1, Y_2 = -1) = \frac{0.8248249}{Z} = 0.18
	$$
	
	$$
	P(Y_1 = 1, Y_2 = 1) = \frac{2.29118}{Z} = 0.5
	$$

	We now repeat the same with domain $\{0, 1\}$ and first compute the potentials for the states $\{(0,0), (0, 1), (1, 0), (1, 1)\}$:
	
	$$ 
	\exp \left \{
	0.251(0) + 0.251(0) + 0.77(0)(0)
	\right \} 
	= 1
	$$
	
	$$ 
	\exp \left \{
	0.251(0) + 0.251(1) + 0.77(0)(1)
	\right \} 
	= 1.285714
	$$
	
	$$ 
	\exp \left \{
	0.251(1) + 0.251(0) + 0.77(1)(0)
	\right \} 
	= 1.285714
	$$
	
	$$ 
	\exp \left \{
	0.251(1) + 0.251(0) + 0.77(1)(0)
	\right \} 
	= 3.571429
	$$
	
	\noindent
	and then the normalization constant 
	
	$$
	Z = 1 + 1.285714 + 1.285714 + 3.571429 = 7.142857
	$$
	
	\noindent
	We divide the  potentials by Z and obtain the probabilities
	
	$$
	P(X_1 = 0, X_2 = 0) = \frac{1}{Z} =  0.14
	$$
	
	$$
	P(X_1 = 0, X_2 = 1) = \frac{1.285714}{Z} = 0.18
	$$
	
	$$
	P(X_1 = 1, X_2 = 0) = \frac{1.285714}{Z} = 0.18
	$$
	
	$$
	P(X_1 = 1, X_2 = 1) = \frac{3.571429}{Z} = 0.5
	$$
	
	We see that both models predict the same probabilities and are therefore statistically equivalent.

	\section{Increasing interaction parameters only changes the marginal probabilities domain in $\{0, 1\}$}\label{A_marginal_p2}

	Here we show that for an Ising model with $p=2$ variables with $\alpha_1, \alpha_2 = 0$ and $\beta_{12} > 0$  it holds that
	
	\begin{equation}\label{claim_1}
	P(X_1=-1) = P(X_2=-1) = P(X_2=1) = P(X_2=1)
	\end{equation}
	
	\noindent
	for the domain $\{-1, 1\}$, and that
	
	\begin{equation}\label{claim_2}
	P(X_1=0) = P(X_2=0) < P(X_1=1) = P(X_2=1)
	\end{equation}
	
	\noindent
	for the domain $\{0, 1\}$.
	
	We first show (\ref{claim_1}). We assume $\alpha_1, \alpha_2 = 0$ and $\beta_{12} > 0$. Then the Ising model is given by 
	
	\begin{align*}
	P(X_1, X_2) &= \frac{1}{Z} \exp\{ \alpha_1 X_1 + \alpha_2 X_2 + \beta_{12} X_2 X_1\} \\
	&=  \frac{1}{Z} \exp\{ \beta_{12} X_2 X_1 \},
	\end{align*}
	
	\noindent
	where $Z$ is the normalizing constant summing over all $2^p=4$ states. We calculate the probability of the four possible states:

	$$
	P(X_1 = 1 , X_2 = -1) = \frac{1}{Z} \exp\{ -\beta_{12} \},
	$$
	$$
	P(X_1 = 1 , X_2 = 1) = \frac{1}{Z} \exp\{ \beta_{12} \},
	$$
	$$
	P(X_1 = -1 , X_2 = -1) = \frac{1}{Z} \exp\{ \beta_{12} \},
	$$
	$$
	P(X_1 = -1 , X_2 = 1) = \frac{1}{Z} \exp\{ -\beta_{12} \}.
	$$
	
	And average over the state of $X_2$ to obtain the marginals probabilities $P(X_1)$:
	
	$$ P(X_1 = 1) = P(X_1 = 1 , X_2 = -1) + P(X_1 = 1 , X_2 = 1) = \frac{1}{Z} \exp\{ -\beta_{12} \} + \frac{1}{Z} \exp\{ \beta_{12} \}$$
	
	$$ P(X_1 = -1) = P(X_1 = -1 , X_2 = -1) + P(X_1 = -1 , X_2 = 1) = \frac{1}{Z} \exp\{ \beta_{12} \} + \frac{1}{Z} \exp\{ -\beta_{12} \}$$
	
	We see that $P(X_1 = 1) = P(X_1 = -1) $. By symmetry the same is true for $X_2$, which proves our claim.\\
	
	We next prove (\ref{claim_2}). We again assume $\alpha_1, \alpha_2 = 0$ and $\beta_{12} > 0$ and calculate the probabilities of the four possible states:
	
	$$
	P(X_1 = 1 , X_2 = 0) = \frac{1}{Z} \exp\{ 0 \},
	$$
	
	$$
	P(X_1 = 1 , X_2 = 1) = \frac{1}{Z} \exp\{ \beta_{12} \},
	$$
	
	$$
	P(X_1 = 0 , X_2 = 0) = \frac{1}{Z} \exp\{ 0 \},
	$$
	
	$$
	P(X_1 = 0 , X_2 = 1) = \frac{1}{Z} \exp\{ 0 \}.
	$$
	
	The marginal probabilities $P(X_1)$ are:
	
	$$ P(X_1 = 1) = P(X_1 = 1 , X_2 = 0) + P(X_1 = 1 , X_2 = 1) = \frac{1}{Z} \exp\{ 0 \} + \frac{1}{Z} \exp\{ \beta_{1,2} \}$$
	
	$$ P(X_1 = 0) = P(X_1 = 0 , X_2 = 0) + P(X_1 = 0 , X_2 = 1) = 2 \frac{1}{Z} \exp\{ 0 \}$$
	
	Since $\exp\{ \beta_{12} \} > \exp\{ 0 \}$, we have $P(X_1 = 1) > P(X_1 = 1) $, if $\beta_{12} >0$. By symmetry the same is true for $X_2$, which proves our claim.\\
	
	Note that if we assume $\beta_{12}  < 0$, (\ref{claim_1}) holds again for $\{-1, 1\}$, while for $\{0, 1\}$ we have 
	
	\begin{equation*}
	P(X_1=0) = P(X_2=0) > P(X_2=1) = P(X_2=1)
	\end{equation*}
	
	\noindent
	instead.

	\section{Derivation of Transformation from $\{0, 1\}$ to $\{-1, 1\}$ and vice versa}\label{sec_IsingTrans}
	
	In this section, we first introduce the Ising model for $p$ variables with domain $\{-1, 1 \}$, which is the domain used in physics applications. Next, we introduce the Ising model for $p$ variables with domain $\{0, 1 \}$, which is mostly used in the statistics literature. We connect both models by deriving a formula of the parameters of one parameterization as a function of the parameters of the other parameterization. This allows us to transform the parameterization based on domain $\{-1, 1 \}$ into the parameterization of domain $\{0, 1 \}$ and vice versa.
	
	In the physics domain, variables can take on values in $\{-1, 1 \}$. The probability distribution of the Ising model for $p$ such random variables is specified by
	
	\begin{equation}\label{eq_Ising_X}
	p(y) = \frac{\exp\left(\sum\limits_{i=1}^p\alpha_iy_i +\sum\limits_{i=1}^{p-1}\sum\limits_{j>i}^p \beta_{ij}y_iy_j\right)}{\sum\limits_y\exp\left(\sum\limits_{i=1}^p\alpha_iy_i +\sum\limits_{i=1}^{p-1}\sum\limits_{j>i}^p \beta_{ij}y_iy_j\right)},
	\end{equation}
	
	\noindent
	where $y, y \in \{-1\text{, }1\}^p$, denotes a configuration of the $p$ random variables, and the sum $\sum\limits_y$ in the denominator denotes a sum that ranges over all $2^p$ possible configurations or realizations of $y$.
	
	From a statistical perspective, the Ising model is a model that is completely determined by the spin variables' main effects and their pairwise interactions. A spin variable in the network tends to have a positive value ($y_i = 1$) when its main effect is positively valued ($\alpha_i > 0$), and tends to have a negative value ($y_i = -1$) when its main effect is negatively valued ($\alpha_i < 0$). Furthermore, any two variables $y_i$ and $y_j$ in the network tend to align their values when their interaction effect is positive ($\beta_{ij} > 0$), and tend to be in different states when their interaction effect is negative ($\beta_{ij} < 0$).
	
	In statistical applications, the Ising model is typically used to describe the probability distribution of $p$ binary random variables,
	
	\begin{equation}
	p(x) = \frac{\exp\left(\sum\limits_{i=1}^p\alpha_i^\ast x_i + \sum\limits_{i=1}^{p-1}\sum\limits_{j>i}^p \beta_{ij}^\ast x_ix_j\right)}{\sum\limits_x\exp\left(\sum\limits_{i=1}^p\alpha_i^\ast x_i + \sum\limits_{i=1}^{p-1}\sum\limits_{j>i}^p \beta_{ij}^\ast x_ix_j\right)},
	\end{equation}
	
	\noindent
	where $x$, $x \in \{0\text{, }1\}^p$, denotes a configuration of the $p$ binary random variables, and again we use $\sum\limits_x$ to denote the sum that ranges over all $2^p$ possible configurations or realizations of $x$.
	
	Even though the model is again completely determined by main effects and pairwise interactions, its interaction parameters $\beta^\ast$ carry a different interpretation than the interaction parameters of the Ising model for variables $Y$ in the $\{-1, 1\}$ domain. Here, two binary variables $x_i$ and $x_j$ in the network tend to both equal one ($x_ix_j = 1$) when their interaction effect is positive ($\beta_{ij}^\ast >0$), but their product tends to equal zero ($x_ix_j = 0$) when their interaction effect is negative ($\beta_{ij}^\ast <0$). That is, whenever the interaction between two binary variables $x_i$ and $x_j$ in the network is negative ($\beta_{ij} < 0$), they tend to be in one of the states $\{0\text{, }0\}$, $\{0\text{, }1\}$ or $\{1\text{, }0\}$.
	
	Despite the different interpretations of the two Ising model formulations, one can traverse the two specifications by a simple change of variables. To wit, assume that we have obtained an Ising model for $p$ binary variables $p(x)$ and wish to express its solution in terms of the variables in the $\{-1, 1 \}$ domain, then we require the change of variables
	
	\begin{equation}
	x_i = \frac{1}{2}(y_i + 1) \text{  with inverse relation  } y_i = 2x_i - 1.
	\end{equation}
	
	We use this transformation in the distribution of the binary random variables,
	
	\begin{align}
	p(x) &= \frac{\exp\left(\sum\limits_{i=1}^p\alpha_i^\ast x_i +\sum\limits_{i=1}^{p-1}\sum\limits_{j>i}^p \beta_{ij}^\ast x_ix_j\right)}{\sum\limits_x\exp\left(\sum\limits_{i=1}^p\alpha_i^\ast x_i + \sum\limits_{i=1}^{p-1}\sum\limits_{j>i}^p \beta_{ij}^\ast x_ix_j\right)} \nonumber \\
	&= \frac{\exp\left(\sum\limits_{i=1}^p\alpha_i^\ast \frac{1}{2}(y_i + 1) +\sum\limits_{i=1}^{p-1}\sum\limits_{j>i}^p \beta_{ij}^\ast \frac{1}{2}(y_i + 1)\frac{1}{2}(y_j + 1)\right)}{\sum\limits_y\exp\left(\sum\limits_{i=1}^p\alpha_i^\ast \frac{1}{2}(y_i + 1) +\sum\limits_{i=1}^{p-1}\sum\limits_{j>i}^p \beta_{ij}^\ast \frac{1}{2}(y_i + 1) \frac{1}{2}(y_j + 1)\right)} = p(y),
	\end{align}
	
	\noindent
	and observe that this transformation affects both main effects and pairwise interactions. Working out the sum over pairs of variables, we find
	
	\begin{align}
	\sum\limits_{i=1}^{p-1}\sum\limits_{j>i}^p \beta_{ij}^\ast \frac{1}{2}(y_i + 1)\frac{1}{2}(y_j + 1) 
	&= \sum\limits_{i=1}^{p-1}\sum\limits_{j>i}^p \frac{1}{4} \beta_{ij}^\ast\left( y_iy_j +  y_i + y_j + 1 \right) \nonumber \\
	&= \sum\limits_{i=1}^{p-1}\sum\limits_{j>i}^p \frac{1}{4} \beta_{ij}^\ast y_iy_j + \sum\limits_{i=1}^{p-1}\sum\limits_{j>i}^p \frac{1}{4} \beta_{ij}^\ast y_i + \sum\limits_{i=1}^{p-1}\sum\limits_{j>i}^p \frac{1}{4} \beta_{ij}^\ast y_j \nonumber \\
	&\qquad + \sum\limits_{i=1}^{p-1}\sum\limits_{j>i}^p \frac{1}{4} \beta_{ij}^\ast \nonumber \\
	&= \sum\limits_{i=1}^{p-1}\sum\limits_{j>i}^p \frac{1}{4} \beta_{ij}^\ast y_iy_j + 
	\sum\limits_{i=1}^p \sum\limits_{\substack{j=1 \\ j \neq i}}^p \frac{1}{4} \beta_{ij}^\ast y_i + \sum\limits_{i=1}^{p-1}\sum\limits_{j>i}^p \frac{1}{4} \beta_{ij}^\ast \nonumber \\
	&= \sum\limits_{i=1}^{p-1}\sum\limits_{j>i}^p \frac{1}{4} \beta_{ij}^\ast y_iy_j + \sum\limits_{i=1}^p \frac{1}{4} \beta_{i+}^\ast y_i + \sum\limits_{i=1}^{p-1}\sum\limits_{j>i}^p \frac{1}{4} \beta_{ij}^\ast \,,
	\end{align}

	\noindent
	where the first term reflects pairwise interactions between the variables $y$, the second term reflects main effects of the  variables with main effect $\beta_{i+}^\ast = \sum\limits_{j = 1}^p \beta_{ij}^\ast$, and the last term is constant with respect to (w.r.t.) the variables $y$. Similarly, we can express the sum over the main effects as
	
	\begin{equation}
	\sum\limits_{i=1}^p\alpha_i^\ast \frac{1}{2}(y_i + 1) = \sum\limits_{i=1}^p\alpha_i^\ast \frac{1}{2}y_i + \sum\limits_{i=1}^p\alpha_i^\ast \frac{1}{2},
	\end{equation}
	
	\noindent
	where the last term is again constant w.r.t. the variables $y$. 
	Collecting the main effects, 
	
	\begin{equation}
	\sum\limits_{i=1}^p\frac{1}{2}\alpha_i^\ast y_i + \sum\limits_{i=1}^p \frac{1}{4} \beta_{i+}^\ast y_i = \sum\limits_{i=1}^p\left(\frac{1}{2}\alpha_i^\ast + \frac{1}{4}\beta_{i+}^\ast\right)y_i,
	\end{equation}
	
	\noindent
	and constant terms,
	
	\begin{equation}
	C = \sum\limits_{i=1}^p \frac{1}{2}\alpha_i^\ast + \sum\limits_{i=1}^p\sum\limits_{j=1}^p \frac{1}{4} \beta_{ij}^\ast,
	\end{equation}
	
	\noindent
	we obtain:
	
	\begin{align}
	p(y) &= \frac{\exp\left(\sum\limits_{i=1}^p\left( \frac{1}{2}\alpha_i^\ast + \frac{1}{4}\beta_{i+}^\ast\right)y_i
		+\sum\limits_{i=1}^{p-1}\sum\limits_{j>i}^p \frac{1}{4} \beta_{ij}^\ast y_iy_j + C\right)}{\sum\limits_y\exp\left(\sum\limits_{i=1}^p\left( \frac{1}{2}\alpha_i^\ast + \frac{1}{4}\beta_{i+}^\ast\right)y_i
		+\sum\limits_{i=1}^{p-1}\sum\limits_{j>i}^p \frac{1}{4} \beta_{ij}^\ast y_iy_j + C\right)}
	\nonumber \\
	&= \frac{\exp\left(\sum\limits_{i=1}^p\left( \frac{1}{2}\alpha_i^\ast + \frac{1}{4}\beta_{i+}^\ast\right)y_i+ \sum\limits_{i=1}^{p-1}\sum\limits_{j>i}^p \frac{1}{4} \beta_{ij}^\ast y_iy_j \right)}{\sum\limits_y\exp\left(\sum\limits_{i=1}^p\left( \frac{1}{2}\alpha_i^\ast + \frac{1}{4}\beta_{i+}^\ast\right)y_i+\sum\limits_{i=1}^{p-1}\sum\limits_{j>i}^p \frac{1}{4} \beta_{ij}^\ast y_iy_j\right)},
	\end{align}
	
	\noindent
	which is equal to the Ising model for variables in the $\{-1, 1 \}$ domain when we write $\alpha_i = \frac{1}{2}\alpha_i^\ast + \frac{1}{4}\beta_{i+}^\ast$ and $\beta_{ij} = \frac{1}{4}\beta_{ij}^\ast$.  In a similar way, one can obtain the parameter values of the binary case from a solution of the Ising model for variables in the $\{-1, 1 \}$ domain using $\alpha_i^\ast = 2\alpha_i - 2 \beta_{i+}$ and $\beta_{ij}^\ast = 4\beta_{ij}$. 
	Thus, we can obtain the binary Ising model parameters $\alpha^\ast$ and $\beta^\ast$ from a simple transformation of the $\{-1, 1\}$ coded Ising model parameters $\alpha$ and $\beta$, and vice versa. Table \ref{tab:tab2} summarizes these transformations:

	\begin{table}[h]
		\centering
		\begin{tabular}{ r | c c }
			\textbf{Transformation} & $\alpha$ & $\beta$\\
			\hline
			$\{0,1\} \Rightarrow \{-1,1\}$ & $\alpha_i = \frac{1}{2}\alpha_i^\ast + \frac{1}{4}\beta_{i+}^\ast$ & $\beta_{ij} = \frac{1}{4}\beta_{ij}^\ast$  \\
			$\{-1,1\} \Rightarrow \{0,1\}$ & $\alpha_i^\ast = 2\alpha_i - 2 \beta_{i+}$ & $\beta_{ij}^\ast = 4\beta_{ij}$ 
		\end{tabular}
		\caption{Transformation functions to obtain the threshold and interaction parameters in one parameterization from the threshold and interaction parameters in the other parameterization. Parameters with asterisk indicate parameters in the $\{0, 1\}$ domain.}
		\label{tab:tab2}
	\end{table}

\section{Model equivalence across domains with penalized estimation}\label{A_penMS}

If one estimates the Ising model with an unbiased estimator, one can estimate with domain $ \{0, 1\}$ and obtain by transformation the estimates one would have obtained by estimating with domain $\{-1,1\}$ (and vice versa). In this section we ask whether this is also the case for penalized estimation, which is a popular way to estimate the Ising model \citep[e.g.][]{van2014new, ravikumar2010high}.

In penalized estimation, the likelihood is maximized with respect to a constraint $c$, typically on the $\ell_1$-norm of the vector of interaction parameters  $\beta_{ij}$

$$
\sum_{i=1}^{p} \sum_{\substack{j=1\\ j \neq i}}^{p} | \beta_{ij} | < c.
$$

Estimation with an $\ell_1$-penalty is attractive because it sets small parameter estimates to zero, which makes it easier to interpret the model. The key problem in this setting is selecting an appropriate constraint $c$. A popular approach is to consider a sequence of candidate constraints $C = \{c_1, \dots, c_k\}$ and select the $c_i$ that minimizes the Extended Bayesian Information Criterion (EBIC) \citep{foygel2010extended}, which extends the BIC \citep{schwarz1978estimating} by an additional penalty (weighted by $\gamma$) for the number of nonzero interaction parameters

$$
\text{EBIC}_{c_i} = -2 LL_{c_i}+ s_0 \log n + 4 s_0 \gamma \log p
,
$$

\noindent
where $LL_{c_i}$ is the maximized log-likelihood under constraint $c_i$, $s_0$ is the number of nonzero interaction parameters, $n$ is the number of observations and $p$ the number of estimated interaction parameters.

We are interested in whether selecting models with this procedure in the two domains, $\{0, 1\}$ and $\{-1, 1\}$, leads to statistically equivalent models.  This is indeed the case for the following reason: assume that $c^*$ minimizes the EBIC for domain $\{0, 1\}$, then from the transformation in Table \ref{tab:tab2_intro}, $\frac{c^*}{4}$ should give the lowest EBIC in domain $\{-1,1\}$, because the constraint $|| \beta^* ||_1 < c^*$ on $\{0, 1\}$ is equivalent to the constraint $|| \beta^* ||_1 < \frac{c^*}{4}$ on $Y$.  Thus, if $\frac{c^*}{4}$ is included in the candidate set $C$, when estimating in domain $\{-1,1\}$, two statistically equivalent models should be selected. Note that exactly $\frac{c}{4}$ has to be included, because a slightly larger/smaller constraint can lead to a very different model, if the number of nonzero parameter changes. This nonlinearity arises from the EBIC, in which $s_0$ decreases by 1 (large change) if some parameter with a tiny value (e.g. 0.0001) is set to zero (small change). Therefore, in order to ensure statistically equivalent models one would need to search a dense sequence $C$. Clearly, this is unfeasible in practice. This means that, in practice $\ell_1$-regularized estimation can return models from domains $\{0, 1\}$ and $\{-1,1\}$ that are not statistically equivalent. We leave the task of investigating this issue for different estimation algorithms for future research. In what follows we provide an extended version of this argument.

We define:

\begin{align*}
	c^* &= \arg_{c \in C} \min \text{EBIC}_c \\
	      &= \arg_{c \in C} \min -2 \log \left [ 
	      \frac{1}{Z} \prod_{m=1}^{n}  \exp \left \{  
	      \sum_{i=1}^p \alpha_i^* X_i + \sum_{i=1}^{p} \sum_{\substack{j=1\\ j \neq i}}^{p} \beta_{ij}^*  X_i X_j
	      \right \}
	      \right] \\
	      &+ s_0 \log n + 4 s_0 \gamma \log [p(p-1) / 2], 
\end{align*} 

\noindent
with constraint

$$
\sum_{i=1}^{p} \sum_{\substack{j=1\\ j \neq i}}^{p} | \beta_{ij}^* | < c,
$$

\noindent
where $s_0$ is the number of nonzero interaction parameters, $n$ is the sample size, $p$ is the number of variables $\frac{p(p-1)}{2}$ is the total number of interaction parameters, and $\gamma$ is a tuning parameter.

Now, we would like to show that if $c^*$ minimizes the EBIC in domain $\{0,1\}$, then $4 c^*$ minimizes the EBIC in $\{-1,1\}$.

We use the transformation in Table \ref{tab:tab2_intro} to rewrite the EBIC into the parameterization implied by $\{-1, 1\}$:


\begin{align*}
c^* &= \arg_{c \in C} \min \text{EBIC}_c \\
&= \arg_{c \in C} \min -2 \log \left [ 
\frac{1}{Z} \prod_{m=1}^{n}  \exp \left \{  
\sum_{i=1}^p (\frac{1}{2} \alpha_i^* + \frac{1}{4} \sum_{\substack{j=1 \\ j\neq i}} \beta_{ij}^*) X_i + \sum_{i=1}^{p} \sum_{\substack{j=1\\ j \neq i}}^{p} \frac{1}{4} \beta_{ij}^*  X_i X_j
\right \}
\right] \\
&+ s_0 \log n + 4 s_0 \gamma \log [p(p-1) / 2], 
\end{align*} 

\noindent
with constraint

$$
\sum_{i=1}^{p} \sum_{\substack{j=1\\ j \neq i}}^{p} | \frac{1}{4} \beta_{ij}^* | < c^*.
$$

We can rewrite the constraint into

$$
\sum_{i=1}^{p} \sum_{\substack{j=1\\ j \neq i}}^{p} | \beta_{ij}^* | < 4c^*.
$$

The last inequality shows that the constraint is $4$ times larger for the parameterization in domain $\{0,1\}$. Or the other way around, the constraint is $\frac{1}{4}$ times smaller in $\{-1,1\}$ compared to $\{0,1\}$.

We know that the models are statistically equivalent across domains. Therefore, the likelihood of the model with constraint $c$ in domain $\{0,1\}$ is equal to the likelihood of the model with constraint $\frac{c}{4}$ in domain $\{-1,1\}$. Now, since the transformation never changes a zero estimate in a nonzero estimate or vice versa with probability 1, also the terms $s_0 \log n + 4 s_0 \gamma \log [p(p-1) / 2]$ in the EBIC remain constant across domains. It follows that, if $c^* = \arg_{c \in C} \min \text{EBIC}_c$ in domain $\{0,1 \}$, then $\frac{c^*}{4} = \arg_{c \in C} \min \text{EBIC}_c$ in domain $\{-1, 1\}$.

\end{document}